\documentclass[%
 reprint,
 superscriptaddress,
 amsmath,amssymb,
 aps,
 pre,
]{revtex4-2}

\usepackage{graphicx}
\usepackage{bm}
\usepackage{xcolor}

\begin{document}

\title{Age-targeted failure in evolving networks with addition and deletion of vertices}

\author{Peter Mann}
\email{peter.mann@diai.io}
\affiliation{Data Insights AI, CodeBase Edinburgh, Argyle House, 3 Lady Lawson Street, Edinburgh, EH3 9DR, United Kingdom}

\author{Simon Dobson}
\affiliation{School of Computer Science, University of St Andrews, St Andrews, Fife KY16 9SX, United Kingdom}

\date{\today}

\begin{abstract}
We present a message passing formalism to study the age-targeted robustness of networks that have evolved under the addition and deletion of vertices. We show that in the growing regime removing the oldest vertices fragments the network far sooner than random failure, while the oldest core is far more robust under targeted removal of the young. At a balanced turnover of vertices the attack value vanishes identically and we prove that the giant component is invariant under arbitrary age-targeting.
\end{abstract}

\maketitle

\emph{Introduction.}---%
Real networks are rarely static. Vertices arrive, acquire connections,
and depart, and the interplay of accrual and loss shapes the structure
that any dynamical process on the network inherits
\cite{albert2002,dorogovtsev2002,holme2012temporal,sarshar2004scalefree,Ben-Naim_Krapivsky_2007,Ghoshal_Newman_2007,budnick2025pard,Budnick_Biham_Katzav_2025}. Moore \textit{et al.}~\cite{Moore_Ghoshal_Newman_2006} introduced an exactly solvable model for addition-deletion networks. In their model a vertex joins the network per unit time with $c$ edges and either uniformly or preferentially chooses vertices to connect to. Vertices are removed uniformly at random at rate $r$, the single parameter that interpolates between pure growth $(r=0)$ and the balanced turnover of a network of constant size $(r=1)$. Moore \textit{et al.} find the stationary degree distribution of this process in closed form for several rules by which the incoming vertex selects its partners. It is known that such a growth process creates networks with non-trivial correlations among the degrees of neighbours. Vertices present early
accumulate connections from later cohorts, so the degrees of neighbours are positively correlated \cite{Krapivsky_Redner_2001,PhysRevE.71.036127}, and remain so under random vertex deletion \cite{garciadomingo2008degree}. These correlations have previously been rationalised via the age of a vertex \cite{saldana2007continuum, garciadomingo2008degree}.

A separate and mature body of work concerns robustness: how
much of a network survives, as a connected whole, when vertices fail \cite{albert2000,PhysRevE.67.026126,Newman_Strogatz_Watts_2001,cohen2001,callaway2000,cohen2000,li2021percolation}. Correlations are not in themselves fatal to the ensemble theory; correlated critical singularities have been classified \cite{goltsev2008percolation} as well as targeted attack \cite{shiraki2010cavity}. It is well known that targeted attack of the high-degree hubs is the most efficient way to dismantle a network. However, most models assume that the correlation is prescribed as an input for the robustness study. A grown network denies exactly that: here the correlation is an output of the dynamics, fixed by the growth model rather than free to be specified.

For addition--deletion networks the correlation is generated by a single coordinate, the age $\sigma$ of a vertex. Growing networks map onto hidden-variable ensembles in which the hidden variable is the arrival time \cite{PhysRevLett.89.258702,PhysRevE.66.066121, PhysRevE.68.036112}. Every edge ties the ages of its two endpoints together through the moment of its formation, so conditioned on age the neighbourhoods of a vertex are independent, and the message passing that solves percolation on locally tree-like graphs
\cite{PhysRevLett.113.208702,10.1098/rspa.2022.0774} applies age by age.

Age is worth resolving because it is frequently the only structural coordinate an attacker might have. The distinction between degree and age is one of cost and disclosure. An attack graded by degree presupposes the edge list, a global object that has to be assembled, kept current, and is often precisely the thing an operator guards. Age requires only one timestamp local to each vertex. An attacker requires less information about the network as a whole when conditioning upon vertex age; only knowledge of when vertices arrive and depart.

Age is also the upstream variable of the pair, since growth makes age generate degree: an arrival time forecasts what a vertex will accumulate, where a present degree records only what it has accumulated, and it is the forecast that a defender needs when choosing which cohort to protect before any damage is done. There is significant prior research on identifying the oldest vertex in a graph. The root of a growing tree can be localised to a set whose size does not grow with the tree \cite{bubeck2017adam,frieze2017looking}, and arrival order in general graphs can be inferred from a single snapshot, subject to a
precision--density tradeoff \cite{navlakha2011archaeology,sreedharan2019inferring} and to a sharp
transition in recoverability as the attachment mechanism varies
\cite{young2019phase}. However, much of that literature requires knowledge of the full edge list.

In this Letter we use message passing to study age-targeted failure of grown networks. Because our message passing resolves the network by age, the retention probability of a vertex may be an arbitrary function $\phi(\sigma)$ of its age. The formulation delivers the giant component under uniform failure, under removal of the oldest fraction of the network, under removal of the youngest, and under any schedule between these extremes.

Three results follow, each confirmed by direct simulation of the
evolving network. In the growing regime the attack hierarchy fans out. Targeting the older vertices outperforms the targeting of younger vertices, other parameters being fixed. Finally, at balanced turnover the hierarchy closes exactly and no age-based schedule $\phi(\sigma)$ outperforms random failure.

\emph{The model and its age structure.}---%
Per unit time one vertex is added with $c$ edges to existing vertices
chosen uniformly at random, and $r$ vertices, selected uniformly, are
deleted together with their edges \cite{Moore_Ghoshal_Newman_2006}. We set $\sigma$ to be the continuous age coordinate and $N$ the current number of vertices; $\sigma$ advances by $d\sigma = 1/N$ per step, so a vertex is deleted at rate $r$ in its own age and survives to age $\sigma$ with probability
$e^{-r\sigma}$. Births occur at rate $N$ per unit $\sigma$, the
population grows as $e^{(1-r)\sigma}$, and a uniformly chosen vertex has
age density
\begin{equation}
    \rho(\sigma) = e^{-\sigma},
    \label{eq:age-density}
\end{equation}
for every $r \le 1$.

A vertex of age $\sigma$ meets its neighbours through two channels,
distinguished by the ages of the vertices at the far ends of its edges.
The $c$ \emph{created} edges were formed at its birth. Each target was
a uniformly chosen vertex at the moment of formation and hence carried
age $w\sim\mathrm{Exp}(1)$ by Eq.~\eqref{eq:age-density}; a coordinate
time $\sigma$ later the target is present with probability
$e^{-r\sigma}$ and, if present, has age $\sigma+w$. The \emph{received}
edges arrive throughout life. Per unit $\sigma$ the $N$ incoming
vertices each direct $c$ edges at uniform targets, creating $Nc$
attachment slots, each of which lands on a given vertex with
probability $1/N$; a vertex therefore receives edges as a Poisson
process of rate $c$ in its own age, independent of $N$. An edge
received when the vertex had age $\sigma-\tau$ was directed by a vertex
then of age zero, so its root has current age $\tau$ and is present
with probability $e^{-r\tau}$. The rate is constant, so conditional on
their count the elapsed ages are independent and uniform on
$[0,\sigma]$. That the $c$ created edges are matched by a reception
rate $c$ is no accident: they are the same attachment slots, viewed
from the two ends of the resulting edges.

The live degree at age $\sigma$ is therefore the sum of a binomial and
a Poisson component, $\mathrm{Bin}(c, e^{-r\sigma})$ created and, the
received process thinned by root survival,
$\mathrm{Poisson}\!\big(\mathcal I(\sigma)\big)$ received with
$\mathcal I(\sigma) = c\int_0^\sigma e^{-r\tau}d\tau
= c(1-e^{-r\sigma})/r$, giving the age-conditional degree generating
function
\begin{equation}
    F(z,\sigma) = \big[1-(1-z)e^{-r\sigma}\big]^{c}\,
    e^{-(1-z)\,\mathcal I(\sigma)},
    \label{eq:F}
\end{equation}
and $g(z)=\int_0^\infty e^{-\sigma}F(z,\sigma)\,d\sigma$, which at $r=1$
reproduces the closed form of Ref.~\cite{Moore_Ghoshal_Newman_2006}.
The mean degree profile
\begin{equation}
    \bar k(\sigma) = c\,e^{-r\sigma} + \frac{c}{r}\big(1-e^{-r\sigma}\big)
    \label{eq:kbar}
\end{equation}
rises from $\bar k(0)=c$ to the saturation $c/r$ of the oldest
vertices, and averages over Eq.~\eqref{eq:age-density} to the known
$\langle k\rangle = 2c/(1+r)$ \cite{Moore_Ghoshal_Newman_2006}. At
$r=1$ the two terms conspire and $\bar k(\sigma)=c$ exactly, for every
age. Balanced turnover erases the correlation between age and degree at
the level of means, though not, as Eq.~\eqref{eq:F} shows, at the level
of composition: the old carry Poissonian degrees where the young carry
freshly formed ones.

\emph{Age-resolved message passing.}---%
Let each vertex of age $\sigma$ be retained independently with an
arbitrary probability $\phi(\sigma)$. Conditioned on age the parts of
the network reached through different edges are independent, and the
graph is locally tree-like, so component sizes add and their generating
functions multiply. We exploit this to build the generating function of
the finite component containing a retained vertex of age $\sigma$ and
to close it on a pair of cavity profiles. Throughout, $\xi$ counts the
vertices of a component.

Let $\theta_s(\sigma)$ be the probability that a retained vertex of age
$\sigma$ belongs to a finite component of $s$ vertices, with generating
function $G(\xi,\sigma)=\sum_{s\ge1}\theta_s(\sigma)\,\xi^{s}$, each
power of $\xi$ counting one vertex. Write $s_a$ for the number of
vertices reached through the $a$-th created edge ($a=1,\dots,c$) and
$t_\rho$ for the number reached through the $\rho$-th received edge,
with $P_C$ and $P_\rho$ their distributions, the received edges forming
a Poisson set $\mathcal R$ with root ages
$\{\tau_\rho\}\subset[0,\sigma]$. The vertex sits in a component of
size $s$ exactly when these contributions sum to $s-1$,
\begin{align}
    \theta_s(\sigma) = \Bigg\langle
      \sum_{\{s_a\}}\sum_{\{t_\rho\}}
      &\prod_{a=1}^{c} P_C(s_a)\!\!\prod_{\rho\in\mathcal R}\!\!P_\rho(t_\rho)
      \nonumber\\[-2pt]
      &\times\,\delta\Big(s-1,\ \sum_a s_a + \sum_\rho t_\rho\Big)
      \Bigg\rangle_{\!\mathcal R},
    \label{eq:theta-sum}
\end{align}
with $\delta$ the Kronecker delta and $\langle\cdot\rangle_{\mathcal R}$
the average over realisations of the received process,
\begin{equation}
    \big\langle X\big\rangle_{\mathcal R}
    = \sum_{n=0}^{\infty} e^{-c\sigma}\frac{(c\sigma)^{n}}{n!}
      \prod_{\rho=1}^{n}\int_0^\sigma\!\frac{d\tau_\rho}{\sigma}\,
      X\big(\{\tau_\rho\}\big),
    \label{eq:poisson-avg}
\end{equation}
the uniform root ages restating the order statistics of the
constant-rate reception established above. The randomness private to
one received edge, namely the presence, retention and onward reach of
its root, is averaged inside $P_\rho$, so $\mathcal R$ is a marked
Poisson process whose marks are integrated out point by point.

Multiplying Eq.~\eqref{eq:theta-sum} by $\xi^{s}$ and summing over
$s\ge1$, the constraint lets the single power of $\xi$ distribute over
the edges, $\xi^{s}=\xi\prod_a\xi^{s_a}\prod_\rho\xi^{t_\rho}$ on the
support of the delta, and the sum over $s$ then removes the delta.
Every convolution collapses into a product of independent single-edge
sums,
\begin{equation}
    G(\xi,\sigma) = \xi\,\mathcal A(\xi,\sigma)^{c}\,
    \mathcal D(\xi,\sigma),
    \label{eq:G-product}
\end{equation}
the $c$ created edges being independent and identically distributed, so
that their single-edge sums supply the $c$-th power of one channel
function $\mathcal A = \sum_{s\ge0}P_C(s)\xi^{s}$, while
$\mathcal D = \langle\prod_\rho\sum_{t\ge0}P_\rho(t)\xi^{t}
\rangle_{\mathcal R}$ resums the Poisson collection.

Closing the channels requires the onward reach of a neighbour with the
shared edge struck from its own factorisation. The roles reverse across
an edge: an edge the focal vertex created is one its older endpoint
received, and an edge it received is one its younger root created.
Write $v_R(\xi,\sigma')$ for the generating function of the branch
grown from a present, retained vertex of age $\sigma'$ entered through
an edge it received, the vertex itself included in the count and the
shared edge excluded, and $v_C(\xi,\sigma')$ for entry through an edge
it created.

\emph{Created edges.} An older neighbour is absent or unretained with
probability $1-\phi(\sigma+w)e^{-r\sigma}$ and then contributes the
empty branch $s=0$, a factor of unity; otherwise it contributes itself
and its onward reach at its own age $\sigma+w$ through the received
cavity, $v_R(\xi,\sigma+w)$. Averaging over the age increment
$w\sim\mathrm{Exp}(1)$ and using $\int_0^\infty e^{-w}dw=1$,
\begin{equation}
    \mathcal A(\xi,\sigma) = 1 - e^{-r\sigma}\!\!\int_0^\infty\!\!\! e^{-w}
      \phi(\sigma{+}w)\big[1-v_R(\xi,\sigma{+}w)\big]dw.
    \label{eq:A-xi}
\end{equation}
The retention factor sits inside the average because the increment $w$
is the neighbour's own age above $\sigma$, and $\phi$ judges it there.

\emph{Received edges.} A received edge of root age $\tau$ carries its
root, present with probability $e^{-r\tau}$ and retained with
probability $\phi(\tau)$; its onward reach is the created cavity
$v_C(\xi,\tau)$, the shared edge being one the root created. The
single-edge sum $\sum_{t\ge0}P_\rho(t)\,\xi^{t}$ of a received edge of
root age $\tau$ is therefore
\begin{equation}
    H_R(\xi,\tau) = 1-\phi(\tau)e^{-r\tau}\big[1-v_C(\xi,\tau)\big].
    \label{eq:HR}
\end{equation}
Inserting $X=\prod_\rho H_R(\xi,\tau_\rho)$ into
Eq.~\eqref{eq:poisson-avg}, the $n$-fold integral factorises into
identical single-age integrals and the sum resums,
\begin{equation}
    \mathcal D(\xi,\sigma)
    = \exp\Big(-c\!\int_0^\sigma\!\big[1-H_R(\xi,\tau)\big]\,d\tau\Big).
    \label{eq:D-xi}
\end{equation}
This is Campbell's theorem in generating-function form: the
point-process analogue of $\langle y^{n}\rangle = e^{-\lambda(1-y)}$
for a Poisson count $n$ of mean $\lambda=c\sigma$, generalised to the
location-dependent $y = H_R(\xi,\tau)$. Differentiating in $\sigma$ and
inserting Eq.~\eqref{eq:HR},
\begin{equation}
    \partial_\sigma \ln\mathcal D(\xi,\sigma)
      = -\,c\,e^{-r\sigma}\phi(\sigma)\big[1-v_C(\xi,\sigma)\big],
    \label{eq:D-volterra}
\end{equation}
with $\mathcal D(\xi,0)=1$, the Volterra form, accumulating the
vertex's recruitment history as it ages.

The cavity profiles close on the channel functions themselves. A vertex
entered through an edge it created retains $c-1$ created edges and its
whole received collection; a vertex entered through an edge it received
retains all $c$ created edges and, because a Poisson process is
unchanged by the removal of one of its points, a received collection
distributed exactly as before. Hence
\begin{align}
    v_C(\xi,\sigma) &= \xi\,\mathcal A(\xi,\sigma)^{c-1}
        \mathcal D(\xi,\sigma), \label{eq:vC}\\
    v_R(\xi,\sigma) &= \xi\,\mathcal A(\xi,\sigma)^{c}\,
        \mathcal D(\xi,\sigma). \label{eq:vR}
\end{align}
Comparison with Eq.~\eqref{eq:G-product} gives
$G(\xi,\sigma)=v_R(\xi,\sigma)$, the same Poisson fact read once more:
a uniformly chosen vertex is statistically identical to one whose
received channel is short one edge. Dividing Eq.~\eqref{eq:vR} by
Eq.~\eqref{eq:vC} gives the algebraic link $v_R=\mathcal A\,v_C$, so
striking a channel is division by its factor, and a single unknown
profile carries the fixed point once Eq.~\eqref{eq:D-volterra} is
adjoined.

The closed system holds at general $\xi$, where $G$ retains the whole
cluster-size distribution; we extract the giant component at the
boundary $\xi=1$ and do not pursue the general problem here. Since $G$
generates the finite-cluster probabilities, $G(1,\sigma)$ is the
probability that the component of a retained vertex of age $\sigma$ is
finite, which in the locally tree-like limit is the probability that
the vertex is not joined to the giant component. The identity $G=v_R$
makes $\mathcal V(\sigma)\equiv v_R(1,\sigma)$ exactly this
probability; write $\mathcal A(\sigma)$ and $\mathcal D(\sigma)$ for
the channel functions at $\xi=1$. Equation~\eqref{eq:vR} becomes
\begin{equation}
    \mathcal V(\sigma) = \mathcal A(\sigma)^{c}\,\mathcal D(\sigma),
    \label{eq:V}
\end{equation}
Eq.~\eqref{eq:A-xi} becomes
\begin{equation}
    \mathcal A(\sigma) = 1 - e^{-r\sigma}\!\int_0^\infty\! e^{-w}\,
    \phi(\sigma+w)\big[1-\mathcal V(\sigma+w)\big]\,dw,
    \label{eq:A}
\end{equation}
and Eq.~\eqref{eq:D-volterra}, with the cavity
$v_C(1,\sigma)=\mathcal V/\mathcal A=\mathcal A^{c-1}\mathcal D$
supplied by Eq.~\eqref{eq:vC}, becomes
\begin{equation}
    \partial_\sigma \ln \mathcal D(\sigma)
    = -\,c\, e^{-r\sigma}\phi(\sigma)
    \big[1-\mathcal A(\sigma)^{c-1}\mathcal D(\sigma)\big],
    \label{eq:D}
\end{equation}
with $\mathcal D(0)=1$. Below the transition the only solution is
$\mathcal V\equiv1$ and every cluster is finite. Above it,
$1-\mathcal V(\sigma)$ is the probability that a retained vertex of age
$\sigma$ lies in the giant component; a uniformly chosen vertex has age
density $e^{-\sigma}$ and is retained with probability $\phi(\sigma)$,
so the giant component occupies the fraction
\begin{equation}
    S = \int_0^\infty e^{-\sigma}\,\phi(\sigma)
    \big[1-\mathcal V(\sigma)\big]\,d\sigma
    \label{eq:S}
\end{equation}
of the pre-removal network. For constant $\phi$
Eqs.~\eqref{eq:V}--\eqref{eq:S} reduce to uniform site percolation on
the evolving network, and for $\phi=1$ to its intact giant component.
The fixed point couples the backward integral~\eqref{eq:A}, reaching to
all ages above $\sigma$, to the forward equation~\eqref{eq:D},
accumulating the recruitment history below it, and is solved by
iteration on the half-line.

The percolation threshold of any protocol follows from linearising
about $\mathcal V\equiv 1$. Write $\varepsilon = 1-\mathcal V$,
$a = 1-\mathcal A$ and $d = 1-\mathcal D$, all small.
Equation~\eqref{eq:A} is already linear in $\varepsilon$; expanding
Eq.~\eqref{eq:V} gives
$1-\varepsilon=(1-a)^{c}(1-d)=1-c\,a-d+O(2)$; and in Eq.~\eqref{eq:D},
$\ln\mathcal D=-d+O(2)$ while
$1-\mathcal A^{c-1}\mathcal D=(c-1)\,a+d+O(2)$, so integrating from
$\mathcal D(0)=1$,
\begin{align}
    a(\sigma) &= e^{-r\sigma}\!\int_0^\infty\! e^{-w}
        \phi(\sigma+w)\,\varepsilon(\sigma+w)\,dw, \nonumber\\
    d(\sigma) &= c\!\int_0^\sigma\! e^{-r\tau}\phi(\tau)
        \big[(c-1)\,a(\tau) + d(\tau)\big]d\tau,
    \label{eq:linear}\\
    \varepsilon(\sigma) &= c\,a(\sigma) + d(\sigma), \nonumber
\end{align}
and the transition sits where the spectral radius of this linear
forward--backward system reaches unity.

Balanced turnover admits two exact statements. At $r=1$ the survival
factor in Eq.~\eqref{eq:A} merges with the kernel,
$e^{-\sigma}e^{-w}=e^{-(\sigma+w)}$, so the substitution $u=\sigma+w$
gives
$\mathcal A(\sigma)=1-\int_\sigma^\infty m(u)[1-\mathcal V(u)]\,du$
with $m(\sigma)=e^{-\sigma}\phi(\sigma)$ the retained age density, and
every remaining appearance of the protocol in Eqs.~\eqref{eq:D}
and~\eqref{eq:S} is the same combination $m(\sigma)\,d\sigma$. The
substitution $t(\sigma)=\int_0^\sigma m(u)\,du$ therefore absorbs
$\phi$ into the time coordinate, and the fixed point depends on the
protocol only through the total retained fraction
$\int_0^\infty m = 1-f$: \emph{at balanced turnover every age-targeted
protocol with the same removed fraction has the same giant component.}
The same substitution collapses the linearised
system~\eqref{eq:linear} to constant coefficients, $a' = -(c\,a+d)$
and $d' = c(c-1)\,a + c\,d$ on $t\in[0,1-f]$ with $a(1-f)=0$ and
$d(0)=0$, whose eigenvalues are $\pm\sqrt{c}$; a nontrivial solution
first fits the boundary conditions at
$1-f_c = \operatorname{artanh}(1/\sqrt{c})/\sqrt{c}$, giving the
universal threshold
\begin{equation}
    f_c(r{=}1) = 1 - \frac{\operatorname{artanh}(1/\sqrt{c})}{\sqrt{c}},
    \label{eq:fc-closed}
\end{equation}
equal to $0.6198\ldots$ at $c=3$ and vanishing where
$\sqrt{c}\tanh\sqrt{c}=1$, at $c\simeq1.44$, below which the intact
balanced network has no giant component.

\begin{figure}[t]
    \centering
    \includegraphics[width=\columnwidth]{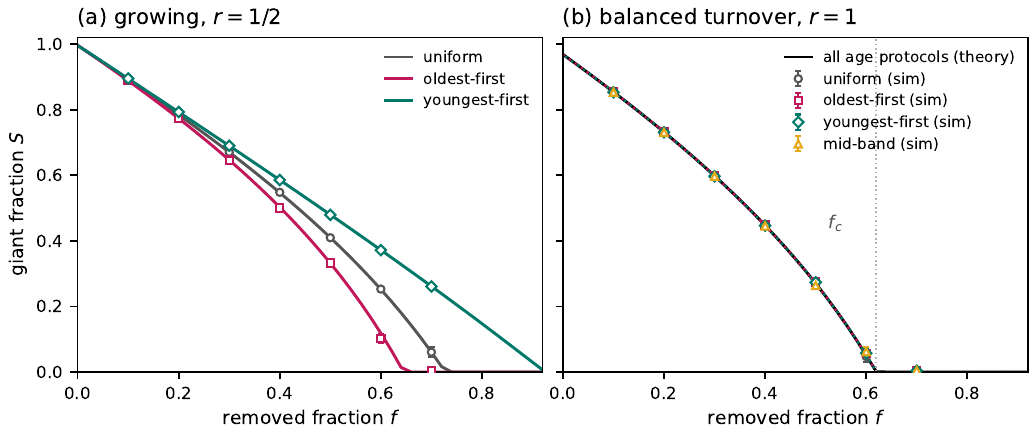}
    \caption{Giant component under age-targeted failure at $c=3$.
    Lines are the exact solution,
    Eqs.~\eqref{eq:V}--\eqref{eq:S}; points are simulations of the
    evolving network.
    (a)~Growing regime, $r=1/2$: the age protocols fan out, oldest-first
    the most damaging and youngest-first the least. (b)~Balanced
    turnover, $r=1$: the exact curves for all removals coincide identically.}
    \label{fig:Sf}
\end{figure}

\emph{Age-targeted failure.}---%
Three protocols span the family. Uniform failure retains every vertex
with the same probability, $\phi(\sigma)=1-f$ with $f$ the removed
fraction. Oldest-first removal deletes every vertex above a cutoff age,
$\phi(\sigma)=\Theta(\sigma^*-\sigma)$, and removes the fraction
$f=e^{-\sigma^*}$ by Eq.~\eqref{eq:age-density}; youngest-first removal
is its complement, $\phi(\sigma)=\Theta(\sigma-\sigma^*)$ with
$f=1-e^{-\sigma^*}$. Figure~\ref{fig:Sf} compares the exact
$S(f)$ curves with simulation of the grown network.

\begin{figure}[t]
    \centering
    \includegraphics[width=0.86\columnwidth]{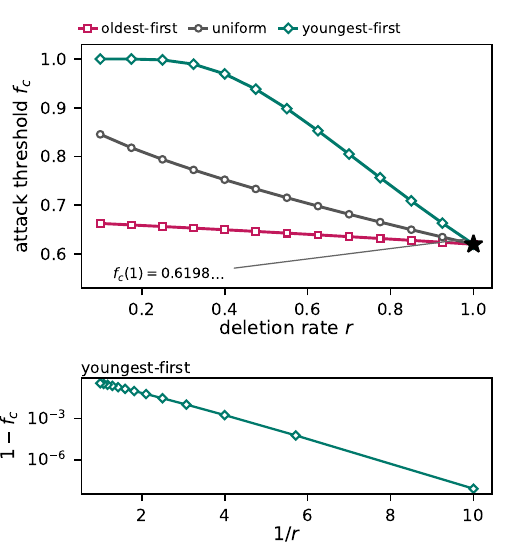}
    \caption{Attack thresholds versus turnover at $c=3$, from the
    linearised fixed point~\eqref{eq:linear}. (a)~The age-protocol fan
    opens as $r$ decreases from balanced turnover, where all protocols
    meet at the universal closed-form value~\eqref{eq:fc-closed}
    (star). (b)~The youngest-first threshold approaches complete
    removal exponentially, $1-f_c \sim e^{-K/r}$ with $K$ slowly
    varying. Lines are theoretical whilst scatter points are the average of Monte Carlo simulation.}
    \label{fig:fc}
\end{figure}

In the growing regime the age fan is wide. At $r=1/2$ oldest-first
removal depresses the giant component below the uniform curve at every
removed fraction, the oldest vertices being the emergent hubs of the
network with degrees saturating at $c/r$, while youngest-first removal
barely dents it [Fig.~\ref{fig:Sf}(a)]. The thresholds quantify the
fan across the whole regime [Fig.~\ref{fig:fc}]: at $r=0.1$ the network
survives random failure up to $f_c\simeq0.845$ but falls to
oldest-first removal already at $f_c\simeq0.663$, and the youngest-first
threshold is indistinguishable from complete removal, $1-f_c$ falling
roughly as $e^{-K/r}$ with $K$ slowly varying
[Fig.~\ref{fig:fc}(b)]. A slowly renewing network is, for practical
purposes, indestructible from its young end: the retained old core is
itself an addition-deletion network whose internal mean degree,
$2c(1-f)^{r}/(1+r)$, stays supercritical until $f$ is exponentially
close to one.

At balanced turnover the fan closes exactly, as the invariance theorem
requires. Panel~(b) in Fig.~\ref{fig:Sf} shows the simulated giant component under uniform, oldest-first, youngest-first and interior-band removal collapsing onto
one universal curve, with the threshold of Eq.~\eqref{eq:fc-closed}
common to all protocols.


\emph{Discussion.}---%
Vertex age is the hidden variable of the addition-deletion model, and
resolving the message passing by age converts the correlations that
obstruct the ensemble theory into the structure that solves it. The
fixed point \eqref{eq:V}--\eqref{eq:S} is exact for an arbitrary
age-dependent retention probability, so a single calculation covers
every failure protocol that reads a vertex's arrival time. In the
growing regime the attack value of the age coordinate is substantial
and asymmetric: the oldest vertices are worth removing and the youngest
are nearly worthless. At balanced turnover it is exactly zero, and from the defender's side is a guarantee. An adversary who removes vertices independently with a
probability depending only on their age leaves precisely the giant
component that random failure of the same size would leave, and fragments the network at precisely the threshold~\eqref{eq:fc-closed}.


\bibliography{ref}

\end{document}